\documentstyle[preprint,tighten,aps,prl]{revtex}
\begin{document}
\draft
\preprint{UTPT-94-10}
\title{Interior Solutions for Non-singular Gravity and the
Dark Star alternative to Black Holes.}
\author{ Neil J. Cornish}
\address{Department of Physics, University of Toronto \\ Toronto,
Ontario M5S 1A7, Canada}
\maketitle
\begin{abstract}
The general equations describing hydrostatic equilibrium are developed for
Non-singular Gravity. A new type of astrophysical structure, a Super Dense
Object (SDO) or ``Dark Star'', is shown to exist beyond Neutron star field
strengths. These structures are intrinsically stable against gravitational
collapse and represent the non-singular alternative to General Relativity's
Black Holes.
\end{abstract}
\pacs{}
\narrowtext

\section{Introduction}
Recently it was discovered that a generalisation of Einstein's theory of
General Relativity (GR) yields an analog of the Schwarzschild
solution which is everywhere non-singular \cite{us}. This remarkable result was
found for a hitherto neglected sector of Moffat's Non-symmetric
Gravity Theory (NGT) \cite{Moff79}. The neglected degrees of freedom
were found to affect drastically the strong field behaviour of the
gravitational field. The smallest, non-vanishing contribution from this
sector removes event horizons and renders the curvature everywhere
finite. This sector of NGT has been dubbed Non-Singular Gravity theory
or NSG in order to distinguish this new physics from the singular
truncation of NGT studied in the literature to date.
The results in Ref.\cite{us} have now been generalised to include the analog of
the Reissner-Nordstrom solution. It was found that the electric
field is everywhere non-singular and both the electromagnetic energy density
and spacetime curvature are everywhere finite\cite{us2}.

These results immediately raise several questions: (1) What do static
structures look like? (2) Can they undergo gravitational collapse?
(3) What is the end result of such a collapse? This paper is primarily
devoted to answering the first of these questions, although tentative
answers are proposed for all three questions. It will be shown that two
distinct classes of interior solution exist. The first class is very similar
to the usual GR solutions, while the second class describes a new and
unusual Super Dense Object (SDO). The GR type structures are unstable
against gravitational collapse, although less so than their GR counterparts.
The SDOs on the other hand are intrinsically stable since their binding
energy tends to zero as they become more compact. Stable SDOs only exist
with radii in a small range near their ``Schwarzschild radius'' $R=2M$.
SDOs with radii much less or much greater than this value are unstable
against {\em gravitational expansion}. For these reasons, it is conjectured
that SDOs represent the non-singular alternative to GR's Black Holes.
Astrophysical phenomena such as Quasi-Stellar Objects (QSOs)
and (Active) Galactic Nuclei (AGNs) may well be SDOs.

The outline of this paper is as follows: First the field equations are
presented. From these the equations describing hydrostatic equilibrium
are derived. The equilibrium equations are then studied in a small region near
the origin, where the existence of two geometrically distinct branches of
solution is established. Each branch of solution is then studied in turn using
a mixture of analytic approximations and numerical surveys. Physical bounds
on the ``skewness parameter'', which controls the departure from GR, are
obtained by considering Neutron stars.

\section{Field Equations}
Einstein's General Theory is based on the assumption that the metric
of spacetime is a symmetric tensor. By removing this assumption, Moffat
was able to construct a non-symmetric theory of gravity where both the
metric and connection are non-symmetric.

The NGT Lagrangian with sources is given by\cite{banff}
\begin{equation}\label{NGT}
{\cal L}=\sqrt{-g}g^{\mu\nu}\left(R_{\mu\nu}(\Gamma)+\frac{2}{3}W_{[\mu,\nu]}
-8\pi T_{\mu\nu}+\frac{8}{3}\pi W_{\mu}S_{\nu} \right)\; ,
\end{equation}
where $R_{\mu\nu}(\Gamma)$ is the generalized Ricci tensor, $W_{\mu}$ is
a Lagrange multiplier, $\Gamma$ refers to the torsion-free connection
$\Gamma^{\lambda}_{\mu\nu}$, $T_{\mu\nu}$ is the energy-momentum tensor
and $S_{\mu}$ is the conserved current which gives rise to NGT charge.
Square brackets denote anti-symmetrisation
and units are chosen so that $G=c=1$ throughout. The field equations that
follow from (\ref{NGT}) are
\begin{eqnarray}
&& g_{\mu\nu,\sigma} - g_{\rho\nu} {\Lambda}^{\rho}_{\mu\sigma} -
g_{\mu\rho} {\Lambda}^{\rho}_{\sigma\nu} = 0 , \\ \nonumber \\
&& {(\sqrt{-g}g^{[\mu \nu]})}_{ , \nu} = 4\pi\sqrt{-g}S^{\mu} , \label{charge}
\\ \nonumber \\
&& R_{\mu \nu}(\Gamma) = \frac{2}{3} W_{[\nu , \mu]}+8\pi (T_{\mu\nu}
-\frac{1}{2}g_{\mu\nu}T) . \label{matter}
\end{eqnarray}
where $\Lambda^{\alpha}_{\beta\gamma}=\Gamma^{\alpha}_{\beta\gamma}
+D^{\alpha}_{\beta\gamma}$, and the tensor $D$ satisfies
\begin{equation}
g_{\rho\nu}D^{\rho}_{\mu\sigma}+g_{\mu\rho}D^{\rho}_{\sigma\nu}
=-\frac{4}{3}\pi S^{\rho}\left(g_{\mu\sigma}g_{\mu\rho}-
g_{\mu\rho}g_{\sigma\nu}+g_{\mu\nu}g_{[\sigma\rho]}\right) \; .
\end{equation}
The generalised Bianchi identities, which result from the
diffeomorphism invariance of NGT, give rise to the matter
response equations
\begin{equation}
{1 \over \sqrt{-g} }\left(g_{\rho\mu}\left(\sqrt{-g}T^{\nu\mu}\right)_{,\nu}
+g_{\mu\rho}\left(\sqrt{-g}T^{\mu\nu}\right)_{,\nu}\right)
+T^{\mu\nu}\left(g_{\mu\rho ,\nu}+g_{\rho\nu ,\mu}
-g_{\mu\nu ,\rho}\right)+\frac{2}{3}W_{[\rho ,\nu]}S^{\nu}=0 \; . \label{cons}
\end{equation}
Taking the matter to be described by a perfect fluid, the energy
momentum tensor is found to be\cite{vin}
\begin{equation}
T_{\mu\nu}=(\rho+p)u_{\mu}u_{\nu}-pg_{\mu\nu} \; ,
\end{equation}
where $u^{\nu}$ is the fluid's four-velocity and $\rho$ and $p$ are the
usual internal energy density and pressure, respectively. The Abelian
gauge invariance of the Lagrangian under $W_{\mu} \rightarrow
W_{\mu}+\lambda_{,\mu}$ ensures the conservation of NGT current density.
This is evidenced in (\ref{charge}) since
\begin{equation}
4\pi(\sqrt{-g}S^{\mu})_{,\mu}=
{(\sqrt{-g}g^{[\mu \nu]})}_{ , \nu\mu} = 0 \; .
\end{equation}
The NGT charge contained in a sphere of radius $r$, in spherical coordinates,
is given by
\begin{equation}
l^2(r)=\int S^{0}\sqrt{-g} dr d\theta d\phi \; .
\end{equation}

The static spherically symmetric interior case was studied for the
truncated version of NGT by Savaria\cite{pierre}. In this work the
pure divergence free parts of $g^{[\mu\nu]}$ were dropped.
Even with this truncation, the analysis was
more complicated than the usual GR case due to the inclusion of
additional source terms arising from the NGT charge. In what follows,
matter will be taken to be NGT charge neutral so that $S^{\mu}=0$
and $l^2(r)=0$. This condition explicitly enforces the strong equivalence
principle.

The most general, static spherically symmetric metric for NGT was found by
Papapetrou\cite{pap} to be
\begin{equation} \label{metric}
g_{\mu\nu}\! = \!\left( \begin{array}{cccc}
\gamma(r)& w(r) & 0 & 0 \\
-w(r) & -\alpha(r) & 0 & 0 \\
0 & 0 & -\beta(r) & f(r)\sin\theta\\
0 & 0 & -f(r)\sin\theta& -\beta(r)\sin^{2}\theta \end{array} \right)\; .
\end{equation}
Taking $l^2(r)$ to be zero demands
$w(r)=0$. The field equations are further simplified by making the
allowed coordinate choice $\beta(r)=r^2$. Both of these simplifications
will be made in the derivation of the hydrostatic equilibrium equations.

\section{Equilibrium Equations}
The field equations can be most compactly expressed in terms of
the following quantities:
\begin{equation}
A=\frac{1}{2}\log(r^4+f^2) \; , \hspace{.4in} B={\rm arctan}(r^2 / f) \; .
\end{equation}
The usual GR expressions are recovered when $A=2\log r$ and $B=\pi /2$.
With $w(r)=0$, the field equation (\ref{charge}) is automatically
satisfied. This is a generic feature of NSG, as the divergence-free
part of $g^{[\mu\nu]}$ {\em is} the NSG sector of NGT. It was
these divergence-free degrees of freedom which were dropped
in previous work on NGT.
The $(tt)$, $(rr)$, $(\theta\theta)$ and $(\theta\phi)$ components
of (\ref{matter}) yield
\begin{eqnarray}
(\log \gamma)''-\frac{1}{2}\left(\log \gamma\right)'\left(\log\left(
{\alpha\over \gamma}\right)\right)'+A'\left(\log \gamma\right)'
&=& 8\pi \alpha\left(\rho +3p\right) \; , \label{tt} \\ \nonumber \\
-2A''+A'(\log \alpha)'-\left((A')^2+(B')^2\right)
-\left(\log \gamma\right)''+\frac{1}{2}\left(\log \gamma\right)'
\left(\log\left({\alpha \over \gamma}\right)\right)'
&=& 8\pi\alpha\left(\rho -p\right) \; , \label{rr} \\ \nonumber \\
1+\left({fB'-r^2 A' \over 2\alpha}\right)'+B'\left({r^2 B'-f A' \over
2\alpha}\right)+\left({fB'-r^2 A' \over 4\alpha}\right)\left(\log
\alpha\gamma \right)'
&=& 4\pi r^2 \left(\rho -p\right) \; , \label{22} \\ \nonumber \\
q+\left({r^2 B'+f A' \over 2\alpha}\right)'-B'\left({fB'-r^2 A' \over
2\alpha}\right)+\left({r^2 B'+f A' \over 4\alpha}\right)\left(\log
\alpha\gamma \right)'
&=& -4\pi f \left(\rho -p\right) \; . \label{23}
\end{eqnarray}
The constant $q$ comes from the Lagrange multiplier field $W_{\phi}
=3q \cos\theta /2 $, and prime denotes derivatives with respect to $r$.
One of these equations may be replaced by the $r$
component of the matter response equation (\ref{cons}), which simplifies
to read
\begin{equation}
p'=-\frac{1}{2}(\rho+p)(\log \gamma)' \; . \label{pp}
\end{equation}
This relation is identical to its GR counterpart. The agreement with GR follows
from NSG obeying the strong equivalence principle when the NGT charge is zero.

The set of equations (\ref{tt})-(\ref{23}) are best studied as three useful
combinations. The first combination is
$f$ times (\ref{22}) plus $r^2$ times (\ref{23}), which yields
\begin{equation}
f+qr^2+\left({(r^4+f^2)B' \over 2\alpha }\right)'-
\left({(r^4+f^2) \over 2\alpha }\right)A'B'+
\left({(r^4+f^2)B' \over 4\alpha }\right)(\log\alpha\gamma)'=0 \; .
\label{comb1}
\end{equation}
The second combination is $f$ times (\ref{23}) minus $r^2$ times (\ref{22}),
which yields
\begin{equation}
fq-r^2+\left({(r^4+f^2)A' \over 2\alpha }\right)'-
\left({(r^4+f^2) \over 2\alpha }\right)(A')^2+
\left({(r^4+f^2)A' \over 4\alpha }\right)(\log\alpha\gamma)'=4\pi(r^4+f^2)
(p-\rho) \; . \label{comb2}
\end{equation}
The third useful combination is (\ref{rr}) plus (\ref{tt}) minus
$4\alpha /(r^4+f^2)$ times (\ref{comb2}) which gives
\begin{equation}
2{r^2-fq \over r^4+f^2}-{1 \over r^4+f^2}\left({(r^4+f^2)A' \over
\alpha}\right)'
+{(A')^2-(B')^2-2A'' \over 2\alpha}=16\pi\rho
\; . \label{comb3}
\end{equation}
These three equations, along with the conservation equation (\ref{pp}),
provide the most useful, complete set of field equations describing
hydrostatic equilibrium in NSG.

Starting with the first of these combinations,
it is straightforward to verify that $f=-qr^2$ solves equation (\ref{comb1}).
With this form for $f$, all the field equations collapse to be identical to
their GR counterparts. Unfortunately, this simple solution is incompatible with
the external Wyman solution\cite{max,us} (the NSG analog of the Schwarzschild
metric) since having $q$ non-zero leads to a jump
discontinuity in the curvature invariant $R_{[\mu\nu]}R^{[\mu\nu]}$ at the
edge of the structure. This follows from the fact that $W_{\phi}=0$ for
the Wyman solution, while $W_{\phi}=3q \cos\theta /2 $ in the interior.
Consequently, the matching of these solutions at the edge of the structure
demands $q=0$.

The complexity of the field equations makes it difficult to combine
them into a simple, fundamental hydrostatic equilibrium
equation. However, considerable insight into the nature of the interior
solutions can be gained by considering power-series solutions about the
origin.
Taking the following expansions:
\begin{eqnarray}
p&=&p_{0}+p_{1}r+p_{2}r^2+ \dots \\
\rho&=& \rho_{0}+\rho_{1}r+\rho_{2}r^2+\dots \\
\gamma&=&\gamma_{0}+\gamma_{1}r+\gamma_{2}r^2+\dots \\
\alpha&=&\alpha_{0}+\alpha_{1}r+\alpha_{2}r^2+\dots \\
f&=&f_{0}+f_{1}r+f_{2}r^2\dots
\end{eqnarray}
and substituting them into (\ref{pp}-\ref{comb3}) leads to
two distinct branches of solution. The first is characterised by having
$\alpha_{0}=1$ and $f_{0}=0$, while the second branch has $\alpha_{0}=0$
and $f_{0} \neq 0$. The first branch is very similar to the usual interior
solutions found in GR, while the second branch is
unlike anything seen before.

Physically, the first branch can be thought of as a matter fields perturbing
Minkowski space, while the second branch looks like matter fields perturbing
the Wyman metric. The existence of this second branch will be seen to
be intimately related to the non-singular nature of the theory. The fact that
there are two branches of solution does not lead to any lack of uniqueness
in the description of a structure since the two branches describe objects
of very different field strengths. For example, the Wymanian branch of
solutions cannot describe the sun, or a white dwarf or even a neutron star.
Instead, this branch describes super dense objects (SDOs), which typically
have radii of $R \leq 2M$. Each branch will now be considered in turn.

\section{Minkowskian Solutions}

The Minkowkian-type solutions have expansions about the origin which
are very similar to their GR counterparts:
\begin{eqnarray}
f&=&\sigma r^3\left(1+\left(\frac{2}{3}\pi\rho_{0}-\frac{2}{5}\pi p_{0}
+\frac{5}{8}\sigma^{2}\right)r^2+\dots \right)\; , \label{fap}\\
p&=&p_{0}-\frac{2}{3}\pi(\rho_{0}+p_{0})(\rho_{0}+3p_{0})r^2+\dots \; , \\
\gamma &=& \gamma_{0}\left(1+\frac{4}{3}\pi\left(\rho_{0}+3p_{0}\right)r^2
+\dots\right) \; , \\
\alpha &=& \left(1-{2{\cal M}(r) \over r}\right)^{-1} \; ,
\end{eqnarray}
where
\begin{equation}
{\cal M}(r)=4\pi\int_{0}^{r}\left[\rho+{21\over 8\pi}\sigma^{2}
+\sigma^{2}\left(\frac{13}{12}\rho_{0}-\frac{11}{4}p_{0}+\frac{5}{16\pi}
\sigma^{2}\right)x^2+\dots \right]x^2 \, dx \; . \label{inmass}
\end{equation}
The above quantity is related to the gravitational mass beneath radius
$r$. The relation is not exact, however, as the matching at the edge of the
structure leads to a complicated relation between ${\cal M}(R)$ and the mass
of the structure $M$. The skew field, $f$, increases the mass of
a structure above that of a structure with a similar matter distribution,
$\rho(r)$, in GR. From (\ref{inmass}) it is clear the mass increase results
from the energy density $\sigma^{2}$ of the skew field.
Since $\sigma^2$ has the dimensions of energy density, it is convenient to
parameterise this quantity as some fraction of the central density:
\begin{equation}
\sigma^2=\epsilon\rho_{0} \; .
\end{equation}
Since the contribution to the effective energy density from $f$ increases
from $21\sigma^2 /4\pi$ with increasing $r$,
it is no surprise that even a small value of $\epsilon$ causes significant
changes. Numerical studies reveal that large departures from the usual
GR results occur once $\epsilon$ reaches $\epsilon_{crit}\sim 0.1 (2M/R)$.
The fact that the critical value for $\epsilon$ scales with the dimensionless
``gravitational potential'' $2M/R$ is in keeping with the general physical
picture of NSG. Large departures from GR occur when the gravitational
potential is large. This relationship is made more precise by considering
the matching of the interior solutions to the exterior Wyman metric.

\subsection{Matching Conditions}
The boundary between the interior and exterior solutions occurs at a
radius $r=R$ defined by $p(R)=0$. The exterior Wyman metric is described by
\begin{eqnarray}
\gamma&=&e^{\nu}\; , \\ \nonumber \\
\alpha&=&{M^2(\nu ')^2 e^{-\nu} (1+s^2) \over (\cosh(a\nu)-\cos(b\nu))^2}\; ,\\
\nonumber \\ \nonumber \\
f&=&{2M^2e^{-\nu}[\sinh(a\nu)\sin(b\nu)+s(1-\cosh(a\nu)\cos(b\nu))] \over
(\cosh(a\nu)-\cos(b\nu))^2} \; ,
\end{eqnarray}
where
\begin{equation}
a=\sqrt{{\sqrt{1+s^2}+1 \over 2}}\; , \hspace{0.5in}
b=\sqrt{{\sqrt{1+s^2}-1 \over 2}}\; ,
\end{equation}
and $\nu$ is given implicitly by the relation:
\begin{equation}\label{impl}
e^{\nu}(\cosh(a\nu)-\cos(b\nu))^2{r^2 \over 2M^2}=\cosh(a\nu)\cos(b\nu)
-1+s\sinh(a\nu)\sin(b\nu) \; .
\end{equation}
The dimensionless quantity $s$ arises as a constant when solving the
vacuum field equations\cite{max}. Unlike the constant of integration $M$, which
is related to a Gaussian integral of a conserved charge, the
constant $s$ is not connected with any conserved quantity in the theory.
Unlike $M$, there is nothing in the theory which could explain why $s$ might
vary from one body to another.
For this reason, it is reasonable to expect that $s$ is a fundamental
constant of nature. This conjecture is supported in the weak field regime where
$s$ is seen to be a coupling constant setting the relative strength of the
skew and symmetric fields. In what follows, $s$ will be taken to
be a universal constant of nature.

Formally matching the solutions at $r=R$ is not very enlightening, since
$\gamma$ is only known as an implicit function of $r$. More can be learnt
by matching onto a small $M/r$ expansion of the Wyman metric, where the
metric functions take the near-Schwarzschild form:
\begin{eqnarray}
\gamma&=&1-{2M \over r}+{s^2 M^5 \over 15 r^5}+{4 s^2 M^6 \over 15 r^6}
        + \dots \; , \\ \nonumber \\
\alpha&=&\left(1-{2M \over r}+{2s^2M^4 \over 9r^4}
+{7s^2M^5 \over 9r^5}+{87s^2M^6 \over 45 r^6} + \dots \right)^{-1}
\; , \\ \nonumber \\
f&=&{sM^2 \over 3}+{2sM^3 \over 3r}+{6sM^4 \over 5r^2} + \dots \; .
\end{eqnarray}
These expansions are only valid for $2M/r \leq 1$, which is not a serious
restriction seeing as most structures have $2M/R <<1$.

Matching $\alpha$ and $f$ at $r=R$ reveals:
\begin{eqnarray}
M&=& M_{0}+{s^2 M_{0}^4 \over 18 R^4}
\left(11M_{0}+
\frac{15}{4}R-\frac{7 \pi}{15}(\rho_{0}+3p_{0})R^3 +\dots \right) \; ,
\label{mass} \\
\nonumber \\
\sigma^2 &=& {s^2 M^4 \over 9 R^6}\left(1+{4M \over R}+\frac{4\pi}{15}
(3p_{0}-5\rho_{0})R^2  +\dots \right) \; , \label{sr}
\end{eqnarray}
where $M_{0}$ is the usual GR expression for the mass
\begin{equation}
M_{0}=M^{{\rm GR}}=4\pi \int_{0}^{R}\rho \, r^2\, dr\; .
\end{equation}
Since the corrections to the GR expression for the mass of the
structure are always positive, the gravitational
mass of a body in NSG will exceed that of a body with a
similar density distribution in GR. This increase in mass is primarily
due to a decrease in the binding energy, which occurs as a result of
the repulsive contributions to the gravitational force coming from the skew
sector. If $s$ is too large, the repulsive contributions from the skew
sector can cause a structure to become unbound. More moderate values of
$s$ serve to stabilise a structure relative to its GR counterpart.
The leading term in (\ref{sr}) can
be used to express $\epsilon$ in terms of $s$:
\begin{equation}
\epsilon = {4\pi \over 27}\left({M \over R}\right)^{3}\left({\bar{\rho}
\over \rho_{0}}\right)\, s^2 \; ,
\end{equation}
where $\bar{\rho}$ is the ``average density'' of the structure,
\begin{equation}
\bar{\rho}={3M \over 4\pi R^3} \; .
\end{equation}
The above expression relating $\epsilon$ and $s$ indicates that the
critical value of $s$ for which large departures from GR begin to be
seen will scale as $s_{crit} \approx R/M$. This indicates that
the most stringent bounds on $s$ will come from the study of Neutron
stars, which is no surprise considering Neutron stars are the only
observationally verified astrophysical objects which require a
relativistic description. The numerical results will show that if
$s$ exceeds $s_{\infty} \sim 20$, Neutron stars become unbound.

One unusual feature of the matching at $r=R$ is that the matching of
the metric functions does not ensure the matching of the gradients
of the metric functions. In GR, any pressure and density profile which
smoothly approaches zero at the edge of the structure leads, via the
field equations, to a smooth matching of $\alpha'$ and $\gamma'$ once
$\alpha$ and $\gamma$ have been matched. This in turn ensures that all
curvature invariants smoothly match at $r=R$, which is the
physically important condition. In NSG the curvature invariants can
smoothly match at the edge of the structure even though there are
jump discontinuities in the gradients of the metric functions, so long
as these jumps obey certain relations which can be derived from the
field equations. For example, the jump in $\gamma'$ is given by the
relation
\begin{equation}
4R^3 \left[{\gamma' \over \gamma}\right]=\left[\left(f'\right)^2\right]\left(
{R^4-f^2 \over R^4+f^2}\right)-\left[f'\right]{8R^3 \over R^4+f^2}-
\left[{\gamma'f' \over \gamma}\right]2f \; , \label{jump}
\end{equation}
where square brackets denote the jump in the enclosed quantity at $r=R$.
In the GR limit, $f=0$ and the above equation yields $[\gamma'\, ]=0$. In
NSG there may or may not be a jump in $\gamma'$, depending on whether
there is a jump in $f'$. For the Minkowskian branch of solutions $f'$ is
always positive in the interior while $f'$ is always
negative in the exterior (if $r>M$), leading to a jump
in $f'$. Equation (\ref{jump}) serves as a useful check on the accuracy
of the numerical solutions, and was found to be satisfied within error
tolerances.

\subsection{Binding Energy}

The internal energy of a body is defined in the usual way to be $E=M-m_{N}N$,
where $m_{N}$ is the rest mass of the $N^{{\rm th}}$ constituent
and $N$ is the conserved particle number for this constituent:
\begin{equation}
N=\int \sqrt{-g}\, J_{N}^{0}\, drd\theta d\phi=\int_{0}^{R}4\pi\,
\sqrt{\alpha(r)(r^4+f(r)^2)}\, n(r)dr \; ,
\end{equation}
and $n(r)$ is the proper number density for this constituent. The total
internal energy can then be broken up into its thermal and gravitational
components as $E=T+V$. The thermal energy is given in terms of the proper
internal material energy  density, $e(r)=\rho(r)-m_{N}n(r)$, by
\begin{equation}
T=\int_{0}^{R}4\pi\, \sqrt{\alpha(r)(r^4+f(r)^2)}\; e(r)dr \; ,
\end{equation}
while the gravitational potential energy $V$ is given by
\begin{equation}
V=M-\int_{0}^{R}4\pi\,\sqrt{\alpha(r)(r^4+f(r)^2)}\; \rho (r)dr \; .\label{v}
\end{equation}
The gravitational binding energy, $\Omega=-V$, is always smaller in NSG
than it would be for the same mass distribution in GR. To leading order, the
binding energy is given by
\begin{eqnarray}
\Omega&=&\Omega^{{\rm GR}}-\frac{15}{8}\sigma^2 R^3- \dots  \nonumber \\
&=&\Omega^{{\rm GR}}-\epsilon\,  M {3 \over 4\pi}\left( {\rho_{0} \over
\bar{\rho}}\right) - \dots \nonumber \\
&=&\Omega^{{\rm GR}}-{15 M s^2 \over 72}\left({M \over R}\right)^3 - \dots \; ,
\label{bind}
\end{eqnarray}
which explains why a large value of $\epsilon$ (or $s$) causes
structures to become unbound. For polytrope equations of state:
\begin{equation}
p=\kappa \rho^{(1+1/n)} \; ,
\end{equation}
the Newtonian expression for the binding energy becomes
\begin{equation}
\Omega^{{\rm Newt.}}={3 \over 5-n}\, {M^2 \over R} \; ,\label{newt}
\end{equation}
which can be combined with (\ref{bind}) to give an expression for
the value of $s$ for which the structure becomes unbound:
\begin{equation}
s_{\infty}=\left({72 \over 5(5-n)}\right)^{{1 \over 2}}\,
\left({R \over M}\right) \; . \label{inf}
\end{equation}
The largest mass White Dwarfs are well approximated by a $n=3$ polytrope,
and have $M/R \approx 4\times 10^{-4}$, which bounds $s$ to be below
$s_{\infty}=6.7\times 10^3$. The major source of error in (\ref{inf}) comes
from the approximate expression for the binding energy, which can be off by
a factor of two for White Dwarfs. The second term in (\ref{bind}) is
essentially exact for White Dwarfs and agrees with the numerical results within
0.1\%. A plot of the dependence of $\Omega /M$ on $s$ is displayed in Fig. 1.
for a White Dwarf with $M/R=2\times 10^{-5}$.

\begin{figure}[h]
\vspace{60mm}
\includegraphics{wdbind.ps}
\vspace{5mm}
\caption{ $(\Omega /M \times 10^5)$ as a function of the skewness constant $s$
for a White Dwarf with central density
$\rho_{0}=1.45\times 10^{-13}\; {\rm km}^{-2}$. The lower line, (a), is the
exact numerical result while the upper line, (b), is the analytic result
employing the Newtonian approximation for the binding energy.}
\end{figure}

The largest mass Neutron stars can be
approximated by a $n=3/2$ polytrope, and have $M/R \approx 0.1$, which
gives the far more stringent bound on $s$ of $s_{\infty}=20$. This value
of $s_{\infty}$ will be roughly $30\%$ too large due to the approximate
value used for the binding energy.
Since Neutron stars are known to exist, and since
they are well described by GR, a tighter bound on $s$ can be obtained by
requiring that the repulsive force causes at most a $10\%$ reduction in
the binding energy. This then gives a maximum allowed value of
$s_{{\rm crit}}=6$. Limiting NSG to be a $1\%$ correction over GR
for Neutron stars gives $s_{{\rm crit}}=2$, and so on.

\begin{figure}[t]
\vspace{65mm}
\includegraphics{fin.ps}
\vspace{5mm}
\caption{$f(r)$ in the interior of a Neutron star of radius $R=0.8322\,
\rho_{c}^{-1/2} = 12.5$ km
when $s=8.9$. The upper line, (a), is an analytic approximation, while the
lower line, (b), is the exact numerical result.}
\end{figure}

Even if $s$ is large enough to cause a $20\%$ reduction in the binding energy,
numerical evaluation shows that the density profile is almost identical to
that in GR, with the difference in $\rho(r)$ never exceeding $0.1\%$ anywhere
in the structure. These studies were done using the equation of state
\begin{equation}
p={\rho_{c} \over 5}\left(\rho \over \rho_{c}\right)^{{5 \over 3}} \; ,
\end{equation}
where $\rho_{c}=1/(72\pi)\; {\rm km}^{-2}$ is a critical density. Above this
density the equation of state goes over to a $n=\infty$ polytrope. Choosing
a central density of $\rho_{0}=0.2 \rho_{c}$ and taking $s=0$ yields a
Neutron star with mass $M=0.0782$, radius $R=0.8325$ and binding energy
$\Omega^{{\rm GR}}=0.0074$ in units of
$\rho_{c}^{-1/2}$. From (\ref{inf}), these values lead to the prediction that
$s=9.7$ will cause a $20\%$ reduction in the binding energy. The numerical
results showed a $20\%$ change was caused by having $s=8.9$, and gave rise to
a
Neutron star with mass $M=0.0798$ and radius $R=0.8322$. Since the graphs of
$\alpha ,\; \gamma ,\; \rho$ and $p$ are essentially the same for both
GR and NSG, the only function worthy of display is the skew field $f$. A
plot of $f$ is displayed in Fig. 2. for the above mentioned Neutron star.
The exact numerical result is shown, along with the power series approximation
given in (\ref{fap}). Even for relativistic situations, such as
Neutron stars, the approximate solution for $f$ gives remarkably good results.

It is interesting to note that the
strongest bounds on the coupling of matter to NGT charge came from Neutron
star calculations, where the NGT charge caused an increase in the binding
energy\cite{cms}. The tightest bound corresponded to a $25\%$ increase in the
binding energy. If both sectors of NGT are taken into account, this bound
is removed since a moderate value of $s$ can cause a compensatory $25\%$
decrease in the binding energy. This possibility is worthy of further
investigation.

If $s$ is a fundamental, dimensionless constant of nature, like
the fine structure constant, it is not
unreasonable to expect $s$ to lie in the range $0.001<s<1$. For
this range of values, {\em NSG has essentially no effect on the description
of any known astrophysical object}, save Black Holes.
Since both the Chandrasekar and Oppenheimer-Volkoff limits for the mass of
White Dwarfs and Neutron stars rely on quasi-equilibrium reasoning, and
since the equilibrium description of these structures is unchanged, the
usual sequence of gravitational collapse will occur. The departure of NSG
from GR comes as the collapse continues, not to a Black Hole, but to
the SDOs to be described in the next section.

\section{Super Dense Objects}

The second branch of interior solution is characterised by having a
rich geometrical structure near the origin. The metric near $r=0$ is of the
form
\begin{eqnarray}
\gamma&=&\gamma_{0}+\gamma_{2} r^2+{\cal O}(r^4) \; ,\\
\alpha&=&\alpha_{2} r^2+{\cal O}(r^4) \; , \\
f&=&f_{0}+f_{2}r^2+{\cal O}(r^4) \; ,
\end{eqnarray}
so that the proper volume scales as $V_{p}\sim r^2$, while two-surfaces
of constant $r$ have surface area $S=4\pi r^2$. The curvature invariants
are all constant near the origin, and are proportional to $(S/V_{p})^2$.
For the vacuum Wyman metric the leading coefficients are given by
\begin{eqnarray}
\gamma_{0}&=&\exp\left[-{\pi \over s}-2-{\pi\, s \over 8}+{\cal O}(s^2)\right]
\; , \\
\alpha_{2}&=&{4\gamma_{0} \over M^2 s^2}\left(1+{\cal O}(s^2)\right) \; , \\
f_{0}&=& M^2\left(4-{s \pi \over 2}+s^2+{\cal O}(s^3) \right) \; ,
\end{eqnarray}
where a small $s$ expansion has been employed. It should be noted that the
limit $f\rightarrow 0$ is achieved by taking $M\rightarrow 0$, {\em not}
$s \rightarrow 0$. The Wymanian interior solutions
possess the same geometry near the origin, although the coefficients in the
expansion are affected by the presence of matter. Interestingly, the ratio
$4\gamma_{0} / \alpha_{2} = s^2 M^2 $ remains almost completely unchanged
in the presence of matter. This serves to highlight a major technical
difficulty
in obtaining numerical solutions to the field equations for this branch of
solution. The standard method for obtaining numerical solutions relies on
power series solutions near $r=0$ to initiate the integration out
from the centre of the structure. This is usually no problem since the region
near $r=0$ is essentially Minkowski space, and no {\it a priori} knowledge of
global parameters, such as the structure's mass, is required. For the Wymanian
branch of solutions this is not the case since the mass of the structure
must be known before it can be determined!

\begin{figure}[h]
\vspace{65mm}
\includegraphics{sdoden1.ps}
\vspace{5mm}
\caption{The density profile $\rho(r)$ for a typical ``neutron SDO''.}
\end{figure}

An obvious solution to this quandary is to break with convention and integrate
from the edge of a structure in toward the centre. This can be successfully
accomplished in GR by specifying a mass and radius and integrating to find
the corresponding central density (although unphysical choices for the mass
and radius will give rise to negative central densities). The success of this
operation in GR is predicated on the fact that smooth matching of $\alpha$
and $\gamma$ automatically results in the smooth matching of all gradients.
In NSG the jump in the gradients must also be specified. For the Minkowskian
branch of solutions non-zero jumps could not be avoided, making this method
difficult to implement. Thankfully, the Minkowskian branch can be treated
using the conventional method. For the Wymanian
branch of solutions $f'$ has the same sign in the interior and
exterior solutions, so it is not imperative that there be a jump in $f'$.
In the absence of a proof that the magnitudes of $f'$ should also match,
it is difficult to proceed. While there is no guarantee that the derivatives
match, there is no doubt that choosing all gradients to match is one
acceptable possibility. This assumption will now be made in order to proceed.

A wide numerical survey was made in order to cover a range of equations of
state and values of $s$. While the details differ, all SDO solutions
studied did fit a standard picture. The generic features include a decrease in
binding energy for radii above or below a central value, $R_{{\Omega}}$,
which typically lay in the range $1.1\, M < R_{{\Omega}} <2.1\, M$.
The central density exhibited similar behaviour, but always peaked at
a slightly smaller value of radius, $R_{{\rho}}< R_{{\Omega}}$,
and dropped more rapidly for larger radii. SDOs with
radii slightly larger than $R_{{\rho}}$ had density profiles which grew
and then dropped, while SDOs with radii below $R_{{\rho}}$ had monotonically
increasing density profiles. A comparison of binding energies at fixed
$\rho_{0}$ for SDOs with radii on either side of $R_{{\rho}}$, revealed that
the SDOs with radii less than $R_{{\rho}}$ were more tightly bound and thus
energetically favoured. This is in agreement with the usual expectation that
stable structures must have monotonically increasing density profiles. Of
course, usual expectations should not be relied upon for these unusual objects.
Varying the value of $s$ did not change this overall picture, although it
did increase $R_{{\Omega}}$ and $R_{{\rho}}$ somewhat, and tended to reduce
peak binding energies (which could reach huge values for small values of $s$).

\begin{figure}[h]
\vspace{65mm}
\includegraphics{sdof1.ps}
\vspace{5mm}
\caption{The skew function $f(r)$ inside a typical ``neutron SDO''.}
\end{figure}

The most important of the above properties is the intrinsic stability of
all SDOs against gravitational collapse. If one tries to crush a SDO
into a smaller radius its binding energy {\em decreases}, and like a
cosmic rubber ball, the SDO will spring back to its equilibrium radius.
This intrinsic stability allows stable SDOs to exist for {\em any mass}.
SDOs represent a non-singular endpoint to gravitational collapse, and provide
nature with a sane alternative to GR's Black Holes. While a SDO resides deep
inside a gravitational well, and can have a surface redshift so large that
it is essentially a ``black star'', the gravitational gradients are
mollified to the extent that matter pressure is able to support the structure.

In order to illustrate these general features, the density profile and
skew field are displayed for a ``neutron SDO''. The equation
of state is designed to roughly model a relativistic core coated with a
crust of non-relativistic neutrons:
\begin{equation}
p=\left\{ \begin{array}{ll}
\frac{1}{3} \rho &\; , \; \rho \geq 1 \\
\frac{1}{5} \rho^{{5 \over 3}} &\; , \; \rho < 1 \; .
\end{array}
\right.
\end{equation}
Here units have been chosen where $\rho_{c}=1$. The density profile shown
in Fig. 3. is for an SDO with $s=0.1$, $M=1$, $R=1.298$. This SDO had
a central density of $\rho_{0}=9.63\times 10^{6}$, and a binding energy of
$\Omega=26.02$. The skew field $f$ is displayed in Fig. 4., and illustrates
the primary difference between the Wymanian and Minkowskian branch of
solutions. The Wymanian branch is characterised by having a large central value
of $f$, with $f_{0} \sim 4 M^2$. The field undergoes a small initial increase
before dropping sharply toward the edge of the structure.
The Minkowskian branch, on the other hand, has $f$ starting from zero and
sharply increasing. Interestingly, the energetically disfavoured SDOs with
radii
above $R_{\rho}$ have skew functions which start near zero and increase
to a peak value before decreasing. This suggest a series of
quasi-equilibrium configurations may exist between the Minkowskian and
Wymanian branch of solutions, although time dependent collapse calculations
will have to be performed to verify this conjecture. Work is currently in
progress\cite{us3} to study the evolution of SDOs through
gravitational collapse in NSG.

The aforementioned neutron SDO has a surface redshift of
$z=2.09\times 10^{5}$, which suggests the name Dark Star is
appropriate for these objects. As $s$ tends to zero, the surface redshifts
associated with SDOs tend to infinity, making Dark Stars as black as
Black Holes.

The neutron SDOs with $s=0.1$ become unbound for radii less than
$R \approx 1.264$. The dependence of binding energy and central density
on radii was studied in detail for the stiff equation of state:
\begin{equation}
p=\rho^{{5 \over 3}}\; .
\end{equation}
This involved finding the density profiles for over fifty different radii
in order to produce plots of $\Omega(R)$ and $\rho_{0}(R)$.
\begin{figure}[h]
\vspace{65mm}
\includegraphics{bind2.ps}
\vspace{15mm}
\caption{The binding energy $\Omega$ as a function of radius $R$ for a
SDO with a stiff equation of state.}
\end{figure}
For most
choices of $s$ and equation of state fewer radii were studied, since it
was sufficient to check that the general picture was the same.
In Fig. 5. the binding energy is plotted against radius for this stiff
equation of state with $s=0.1$. A similar plot of the
central density versus $R$ for these SDOs is displayed in Fig. 6.
For radii beyond $R=2.04$ a new complication enters the picture since the
central density is zero and there is a region of empty space near the origin.
This would require an inner and outer matching to the vacuum Wyman solution,
although this possibility was not pursued since such annular distributions
become unbound.

\begin{figure}[h]
\vspace{60mm}
\includegraphics{cenden.ps}
\vspace{15mm}
\caption{The central density $\rho_{0}$ as a function of radius $R$ for a
SDO with a stiff equation of state.}
\end{figure}

The above description of SDOs is at best a rough sketch, based on a wide, but
by no means exhaustive, numerical survey. Since little is known about the
behaviour of matter for the kinds of densities being considered for SDOs, it
is difficult to choose a sensible equation of state. On the positive side,
the generic features described for SDOs do not seem to be greatly affected
by the choice of equation of state. The possibility that the gradients of
$\alpha\; , \gamma$ and $f$ exhibit jumps at the edge of some SDOs remains to
be investigated. It is possible that a proof can be found to show the
Wymanian branch of solutions never exhibit jumps in metric gradients. No
work has yet been done to study the feasibility of describing QSOs or AGNs
in terms of SDOs, although the prospects are very good. In short,
much work remains to be done to understand these fascinating objects.

\section{Discussion}

The non-singular gravitational theory described in this paper
reproduces every prediction of GR for weak to moderate gravitational fields
to any desired accuracy (taking $s$ small enough). Any experimental prediction
based on systems with $2M/R <1$ will fail to tell NSG and GR apart, unless
$s$ is larger than $\sim 1$. If $s$ is larger than this value there is some
hope that detailed studies of Neutron stars may tell the two theories apart.

The only unique signal for any value of $s$ would come for the super
dense objects believed to exist at the centre of most galaxies. If these
galactic nuclei have masses above $10^{8}M_{\odot}$, GR predicts that
a star passing by will be torn apart by tidal forces once it is within the
horizon of the massive Black Hole, so the star would vanish without
a trace. If the galactic nuclei was a SDO on the other hand, the star's
death would be visible\cite{st}. Unfortunately, if $s$ is too small, the
radiation given off as the star is torn apart would be so heavily redshifted
that no signal could be seen above background.

In the absence of experimental tests to tell NSG and GR apart, there is
only the following aesthetic distinction:
Any non-zero value of $s$ leads to a
non-singular endpoint to gravitational collapse - a Super Dense Object
or Dark Star. In contrast,
GR is always fated to have a singular endpoint to collapse - a Black Hole.
Why cling to the concept of a Black Hole when the smallest imaginable
change to GR banishes these objects from our picture of the universe?

Infinities belong in mathematics, not nature.

\section*{Acknowledgments}
I am grateful for the support provided by a Canadian Commonwealth
Scholarship. I thank Norm Frankel, Janna Levin, John Moffat and Pierre Savaria
for their interest in this work and for carefully reading the manuscript. I
would also like to thank Dick Bond and Glenn Starkmann for several useful
discussions concerning NSG.


\begin{references}
\bibitem{us} Cornish N. J., Moffat J. W., {\em University of Toronto
preprint} UTPT-94-04, gr-qc/9403013, 1994.
\bibitem{Moff79} Moffat J. W., {\em Phys. Rev.} D {\bf 19}, 3554,
1979.
\bibitem{us2} Cornish N. J., Moffat J. W., {\em University of Toronto
preprint} UTPT-94-08, 1994.
\bibitem{banff} For a review of early work on NGT see Moffat J. W. in
{\em Gravitation - A Banff Summer Institute}, eds. Mann R. B. \&
Wesson P., World Scientific, Singapore, p. 523, 1991.
\bibitem{vin} Vincent D., {\em Class. Quant. Grav.} {\bf 2}, 409, 1985.
\bibitem{pierre} Savaria P., {\em Class. Quant. Grav.} {\bf 6}, 1003, 1989.
\bibitem{pap} Papapetrou A., {\em Proc. Roy. Irish Acad. Sci.} {\bf A552}, 69
1948.
\bibitem{max} Wyman M., {\em Can. J. Math.} {\bf 2}, 427, 1950.
\bibitem{cms} Campbell L. M., Moffat J. W. and Savaria P., {\em Ap. J.}
{\bf 372}, 241, 1991.
\bibitem{us3} Cornish N. J., Moffat J. W., {\em University of Toronto
preprint} UTPT-94-11 (in preparation), 1994.
\bibitem{st} I thank Scott Tremaine for this suggestion.
\end{references}
\end{document}